\documentstyle[epsf,aps]{revtex}
\newcommand{\Tr}{\textrm{Tr}}
\newcommand{\sq}{\sqrt}
\tighten
\begin{document}
\title{Improving the Fidelity of Quantum Cloning by Fast Cycling away the Unwanted Transition}
\author{Shubhrangshu Dasgupta and G. S. Agarwal}
\address{Physical Research Laboratory, Navrangpura, Ahmedabad - 380 009, India}
\date{\today}
\maketitle

\begin{abstract}

The fidelity of quantum cloning is very often limited by the accompanying 
unwanted transitions. We show how the fidelity can be improved by using a 
coherent field to cycle away the unwanted transitions. We demonstrate this
explicitly in the context of the model of Simon {\it et al.\/} [J. Mod. Opt.
{\bf 47}, 233 (2000) ; Phys. Rev. Lett. {\bf 84}, 2993 (2000)]. We also 
investigate the effects of the number of atoms on the quality of quantum 
cloning. We show that the universality of the scheme can be maintained by 
choosing the cycling field according to the input state of the qubit.

\end{abstract}


\section{INTRODUCTION}
Quantum no-cloning theorem states that it is impossible to clone perfectly an 
arbitrary {\it unknown\/} pure quantum state \cite{zurek}, or a mixed quantum 
state \cite{mixed}. The origin of this  theorem can be traced to the linearity 
of quantum mechanics.  However imperfect quantum cloning is possible. An optimal
 $1\rightarrow 2$ imperfect cloner has been proposed \cite{buzek}, which is 
universal for all input qubits  and is compatible with no-signalling constraints
 too \cite{gisin1}.  It has been generalized  \cite{gisin2,werner} for the case 
of $N\rightarrow M$ $(M\ge N)$ cloner. The upper bound for the fidelity of a 
$N\rightarrow M$ cloner has been established \cite{bruss}. Results are also
 available for  the optimal cloning of arbitrary pure and mixed states in 
$d$-dimensions $(d \ge 2)$ \cite{werner,ddim}. Besides, cloning of entangled 
states \cite{entang} and of Gaussian-distributed  quantum variables  
\cite{gauss} has been considered. Quantum cloning was originally discussed in 
terms of polarization states of photons \cite{zurek,milonni,mandel}. Making use 
of two two-level atoms with orthogonal transition dipole moments, Mandel 
\cite{mandel} proved that one can make the output of the photon amplifier 
independent of the input polarization.  Perfect cloning of photonic states has 
been proved impossible due to inevitable coexistence of spontaneous emission 
with stimulated emission process \cite{milonni,mandel}. Simon and  co-workers 
\cite{simon1,simon1a} have proposed a quantum cloning machine (QCM) consisting
of three-level atoms.  They have shown that the quantum cloning of a single 
input qubit (polarization state of a photon) using a V-system is possible with a
fidelity $5/6$ at least for shorter interaction times. This value is  optimal 
for a $1 \rightarrow 2$ quantum cloner \cite{buzek}. Similar results for 
$\Lambda$-systems have been reported \cite{simon2}.  Finally note that the 
state-dependent cloning has also been studied extensively \cite{state}.  In this
 paper, we address the question if the fidelity of a V-system cloner can be 
improved by using some type of external field.

The organization of this paper is as follows. In Sec. II, we briefly outline the
 cloning scheme introduced by Simon {\it et al.\/}. We  prove the universality 
of their scheme by using a new basis for the states of the radiation fields. 
 In Sec. III, we examine the reason for imperfect fidelity  in a
V-scheme and introduce a way to improve this fidelity by using a coherent
field. The external field cycles away the unwanted transition responsible for
spontaneous emission. We present both analytical and numerical results for  
the fidelity. We discuss the question of the universality of the scheme.
 In Sec. IV, we examine the question of improvement in fidelity by 
considering a cloner consisting of two V-systems. In Sec. V, we demonstrate
the improvement in average fidelity for the case of a fixed cycling field for 
all input states of the qubit.

\section{QUANTUM CLONING BASED ON STIMULATED EMISSION IN A V-SYSTEM}
\subsection{Optimal Photon Cloner with a V-configuration}
Recently Simon {\it et al.\/} \cite{simon1,simon1a} have proposed a new scheme
 for quantum cloning of a photonic qubit. They considered a cloning device 
consisting of an ensemble of atoms trapped inside a cavity. The relevant atomic 
transitions correspond to the V-system. These are three-level systems with two 
degenerate excited states $|e_1\rangle$ and $|e_2\rangle$ and a common ground 
level $|g\rangle$. The ground level is coupled to the excited states by two
orthogonal field modes $a_1$ and $a_2$, respectively.

In the interaction picture \cite{sakur}, the effective Hamiltonian under dipole
 and rotating wave approximations \cite{allen,scully} can be written as 
\begin{equation}
\label{ezeiham}
H_I= \hbar g \sum_{k=1}^N (\sigma_{+1}^k a_1 +\sigma_{+2}^k a_2) +\textrm{ H. c.}, 
\end{equation} 
where $g$ is the coupling constant between the field-modes and the atoms, 
$\sigma_{+1(2)}^k$'s $[=(|e_{1(2)}\rangle\langle g|)_k]$ are  the raising 
operators between the corresponding states of the $k$-th atom. Here $g$ is 
assumed to be equal for all the atoms  \cite{space}.  Also both the cavity-modes
 are assumed to be resonant with the corresponding atomic transitions.

Let each atom be prepared initially in a mixed excited state  
\begin{equation}
\label{init_mix}
 \rho= \frac{1}{2}(|e_1\rangle\langle e_1| +|e_2\rangle\langle e_2|)
\end{equation}
and the photonic qubit be prepared in  a state $a_1^{\dagger}|0,0\rangle 
\equiv |1,0\rangle $. The time-development operator $U= e^{-i H_I t}$ will 
provide the time-evolution of the entire (atom+photon) system and this is 
used to study quantum cloning. 

The quality of cloning is characterized in terms of fidelity \cite{fdlt}, 
which can be defined \cite{simon1} in the following way:%different ways \cite{simon1}: 
%\begin{mathletters}
\begin{equation}
%F = \sum_{k+l\: \ge\: 2} p'(k,l) \left(\frac{k}{k+l}\right) ~~~;~~~ 
F = \sum_{k=0}^{N+1} \sum_{l=0}^N p(k,l) \left(\frac{k}{k+l}\right).
\end{equation}
%where 
%\begin{equation}
% p'(k,l) =\frac{p(k,l)}{1-p(1,0)-p(0,1)}.
%\end{equation}
%\end{mathletters}
Here $p(k,l)$ represents the probability of finding $k$ photons in the initial 
mode and $l$ photons in the orthogonal mode $a_2$ in the evolved state. It 
should be noted that for an ensemble of $N$ atoms, the maximum value of $k$ will
 be $N+1$, which corresponds to all the atoms decaying to the ground state 
through the emission of the $a_1$-photon. Thus $F$ is a kind of an average of 
the relative frequency of photons in initial mode $a_1$ in the final state.% The
%final time-evolved entangled state contains the field components with various 
%combination of photon numbers in two modes. The difference between the two 
%definitions arises whether we include in the final state those states containing
% less than two photons. The definition $F$ rejects those final states which 
%contains less than two photons. 
 
As shown by Simon {\it et al.\/}, the fidelity is optimal for short interaction
times and for $N=6$. It decreases for later times. They have explained this 
behavior in terms of stimulated and spontaneous emissions on the transitions 
$|e_1\rangle \rightarrow |g\rangle$ and $|e_2\rangle \rightarrow |g\rangle$, 
respectively. If there is an extra photon in $a_1$-mode, it can be  considered 
as a clone of the initial qubit. Note that the probability to get a clone is 
reduced if there is an extra photon in the other ($a_2$) mode, which is due to  
spontaneous emission.

\subsection{Question of Universality of Cloning by a V-system}   

It is mentioned in Ref. \cite{simon1a} that the above scheme is universal, i.e.,
 the V-system cloner can clone even any arbitrary photonic qubit, say, 
$(\alpha a_1^{\dagger} +\beta a_2^{\dagger})|0,0\rangle$ with the same non-unity
 fidelity. Simon {\it et al.\/} argued  that this is because the initial mixed 
state and the Hamiltonian  are invariant under a unitary transformation.
In what follows, we demonstrate explicitly this universality by changing the 
basis to a general qubit state.  

Consider a single atom in V-configuration, initially prepared in an incoherent 
superposition of the two excited states. We will work with wavefuntions and 
hence we use an initial state  
\begin{equation}
\label{einitgen}
|s\rangle = \frac{1}{\sq{2}} \left(|e_1\rangle +e^{i\theta}|e_2\rangle\right).
\end{equation}
and average the final results over all possible values of the parameter $\theta$.  Let the photon be in a superposition state  
\begin{equation}
\label{einitgen1}
b_1^{\dagger}|0,0\rangle  \equiv \left(\alpha a_1^{\dagger} +\beta a_2^{\dagger}\right) |0,0\rangle \equiv \alpha |1,0\rangle +\beta |0,1\rangle
 \end{equation}
as well, where $\alpha$ and $\beta$ are the complex numbers satisfying $|\alpha|^2+|\beta|^2=1$. Note that an average over $\theta$ will give an initial state of the atom, 
which is a mixed state.
We consider the  photon as a qubit \cite{fidel}, which can be in any 
linear superposition of the two orthogonal states. Let us define the
basis state $b_2^{\dagger} |0,0\rangle $, which is orthogonal to (\ref{einitgen1}). The new operators $b_1$ and $b_2$ must satisfy the commutation relations
\begin{equation}
\label{boper}
 [b_1,b_2]=[b_1,b_2^{\dagger}]=0.
\end{equation}
Using Eqs. (\ref{einitgen1}) and (\ref{boper}), we get  
\begin{equation}
b_2^{\dagger} \equiv -\beta^* a_1^{\dagger}+\alpha^* a_2^{\dagger}.
\end{equation}
The time-evolution of the entire system is determined by the Schr\" odinger equation
\begin{equation}
\label{egenschr}
i\hbar \frac{d}{dt}|\Psi (t)\rangle = H_I |\Psi (t)\rangle,  
\end{equation}
where $H_I$ is given by Eq. (\ref{ezeiham}) for $N=1$. We expand $|\Psi(t)\rangle$ in terms of the relevant basis states.
Starting with the initial conditions [Eqs. (\ref{einitgen}) and (\ref{einitgen1})], these  relevant states were found to be
\begin{equation}
 |e_1\rangle |1,0\rangle~;~|e_2\rangle |1,0\rangle~;~|g\rangle |2,0\rangle~;~
|g\rangle|1,1\rangle~;~|g\rangle |0,2\rangle~;~|e_1\rangle |0,1\rangle~;~|e_2\rangle |0,1\rangle.
\end{equation}
The only non-zero expansion amplitudes $C_{\alpha}^{mn}$ at time $t=0$ are 
\begin{equation}
C_{e_1}^{10}(0)=\frac{\alpha}{\sq{2}}~;~C_{e_1}^{01}(0)=\frac{\beta}{\sq{2}}~;~C_{e_2}^{10}(0)=\frac{\alpha}{\sq{2}}e^{i\theta}~;~C_{e_2}^{01}(0)=\frac{\beta}{\sq{2}}e^{i\theta},
\end{equation}
where the subscript (superscript) denotes the atom (photons) in the state $\alpha$ ($m,n$).
Then all the expansion amplitudes can be evaluated in closed form with the following results:
\begin{eqnarray}
\label{esolnb}
C_{e_1}^{10}(t) & = & \frac{\alpha}{\sq{2}} \cos{\left(\sq{2}gt\right)} \nonumber ,\\
C_g^{20}(t) & = & -i\frac{\alpha}{\sq{2}}\sin{\left(\sq{2}gt\right)} \nonumber , \\
C_{e_2}^{10}(t) & = & \frac{1}{2\sq{2}}\left[\left(\alpha e^{i \theta}-\beta\right)+\left(\beta +\alpha e^{i\theta}\right)\cos{\left(\sq{2}gt\right)}\right] \nonumber , \\
C_{e_1}^{01}(t) & = & \frac{1}{2\sq{2}}\left[\left(\beta -\alpha e^{i\theta}\right) +\left(\beta +\alpha e^{i\theta}\right)\cos{\left(\sq{2}gt\right)}\right], \\
C_g^{11}(t) & = & -\frac{i}{2}\left(\beta +\alpha e^{i\theta}\right)\sin{\left(\sq{2}gt\right)} \nonumber, \\
C_g^{02}(t) & = & -i\frac{\beta}{\sq{2}} e^{i\theta} \sin{\left(\sq{2}gt\right)} \nonumber ,\\
C_{e_2}^{01}(t) & = & \frac{\beta}{\sq{2}} e^{i \theta}\cos{\left(\sq{2}gt\right)} \nonumber .
\end{eqnarray}
The reduced density matrix of the field is defined by 
\begin{equation}
\label{ereden}
\rho_F =\Tr_A(|\Psi (t)\rangle \langle \Psi (t)|).
\end{equation}
Using Eq. (\ref{ereden}), the probability $\tilde{p}(k,l)$ that 
$k$ photons will be in $b_1$-mode and $l$ photons in 
$b_2$-mode can be written in terms of $b$-operators as
\begin{equation}
\label{eprobb}
\tilde{p}(k,l)  =  \langle 0,0|\frac{b_1^k\:b_2^l}{\sq{k!\:l!}}\:\rho_F\:\frac{b_1^{\dagger ^k}\:b_2^{\dagger ^l}}{\sq{k!\:l!}}|0,0\rangle .
\end{equation}
Further in order to get the initial atomic state used by Simon {\it et al.\/}, we average $\tilde{p}(k,l)$ over all values of $\theta$.
A lengthy derivation yields the following:
\begin{mathletters}
\begin{eqnarray}
\tilde{p}_a(2,0) & = &\frac{1}{2}\sin^2{\left(\sq{2}gt\right)}, \\
\tilde{p}_a(1,1) & = & \frac{1}{4}\sin^2{\left(\sq{2}gt\right)}, \\
\tilde{p}_a(0,1) & = & \frac{1}{8}\cos^2{\left(\sq{2}gt\right)}-\frac{1}{4}\cos{\left(\sq{2}gt\right)}+\frac{1}{8}, \\
\tilde{p}_a(1,0) & = & \frac{5}{8}\cos^2{\left(\sq{2}gt\right)}+\frac{1}{4}\cos{\left(\sq{2}gt\right)}+\frac{1}{8},
\end{eqnarray}
\end{mathletters}
where $\tilde{p}_a(k,l)$ is the $\theta$-averaged value of $\tilde{p}(k,l)$.
The Eqs. (14) lead to the following expression for the fidelity:
\begin{equation}
%F = \frac{\tilde{p}(2,0)+\frac{1}{2}\tilde{p}(1,1)}{\tilde{p}(2,0)+\tilde{p}(1,1)} = \frac{5}{6}~~~~;~~~~ 
F = \tilde{p}_a(1,0)+\tilde{p}_a(2,0)+\frac{1}{2}\tilde{p}_a(1,1) = \frac{3}{4}+\frac{1}{4}\cos{(\sq{2}gt)}.
\end{equation}
Clearly the fidelity does not depend upon $\alpha$ and $\beta$. This
reflects the fact that the V-scheme is universal as a cloner, which can clone 
even a general superposition of two orthogonal modes of the field,  albeit
imperfectly. A similar result has been reported recently using a different 
method \cite{simon2}.

\section{A METHOD TO IMPROVE THE FIDELITY OF THE V-SCHEME FOR ARBITRARY STATE OF THE INPUT PHOTON}

It is clear that the fidelity of cloning is degraded by the emission of a photon
 of the ``wrong" type. Thus to improve the fidelity one should reduce the 
probability $p(1,1)$ and $p(0,1)$ of emitting in the mode $a_2$, which is caused by the 
atomic population in $|e_2\rangle$-level. One possible way of doing this is to 
{\it cycle\/} this population away, so that this unwanted spontaneous decay does
 not occur very often. We show that it can be done by applying a classical pump 
field, which causes the population in the state $|e_2\rangle$ to pulsate
between a metastable state $|f\rangle$ and $|e_2\rangle$. However any biasing by
the external field is likely to take away the system from universality. We {\it
 get over the problem} by making the bias dependent on the state to be cloned. 
The scheme would then become {\it near universal}.

\begin{figure}
\epsfxsize 10cm
{\center
\epsfbox{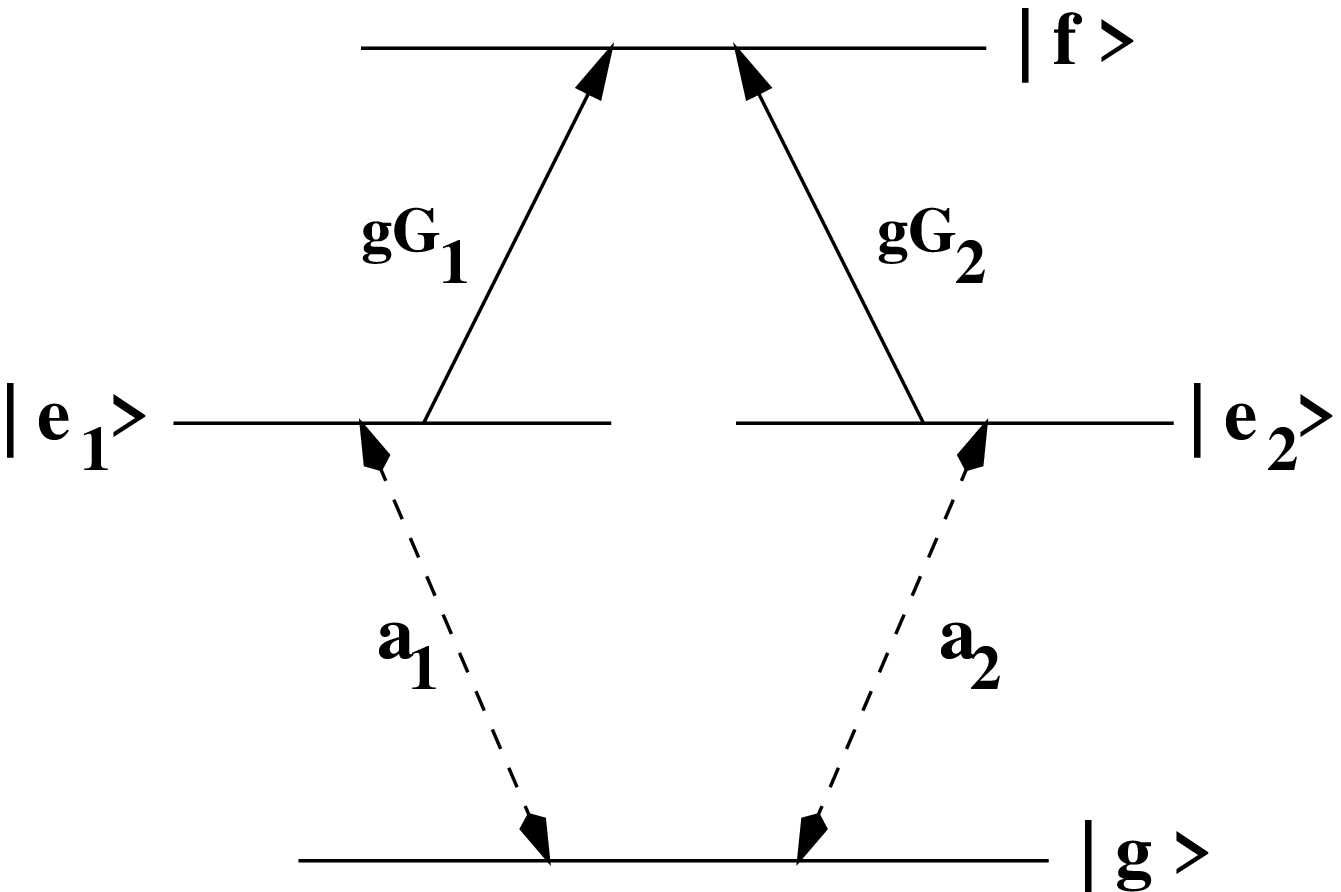}}
\caption{An atomic level configuration to improve the fidelity of quantum 
cloning. Here the classical fields with Rabi frequency $2gG_i~(i=1,2)$ couple 
the levels $|e_1\rangle$ and $|e_2\rangle$ with the metastable state 
$|f\rangle$. The atom is inside a cavity, which allows only two field-modes 
$a_1$ and $a_2$.}
\label{config}
\end{figure}

Consider a four-level atomic configuration as shown in  Fig.\ \ref{config}.  The
 excited states $|e_1\rangle$ and $|e_2\rangle$ are coupled to the common ground
 level $|g\rangle$ through the two orthogonal modes $a_1$ and $a_2$ of the 
quantized electromagnetic field, respectively. The coupling constant between 
each of these excited states and $|g\rangle$ is $g$. We consider the action of 
 classical fields coupling the state $|e_i\rangle~(i=1,2)$ with the metastable 
state $|f\rangle$. The corresponding  Rabi frequency is $2gG_i$, where $G_i$ is
 a multiplying factor and is related to the number of photons in the classical 
field. We assume all the fields to be resonant with the corresponding atomic 
transitions. 

We start with a single atom prepared in the state $|s\rangle$ [Eq. (\ref{einitgen})].
 The initial photonic qubit is in $b_1$-mode. We will work in the interaction 
picture.  Then using the rotating wave approximation to eliminate the 
fast-oscillating energy non-conserving terms, we obtain the effective 
Hamiltonian as  
\begin{equation}
\label{hamilto}
 H_I=\hbar g \left[ \sigma_{+1}a_1 +\sigma_{+2}a_2 + \sum_{i=1,2} G_i|e_i\rangle \langle f |\right] +\textrm{H.c.},
\end{equation}
where $\sigma_{+1(2)}=|e_{1(2)}\rangle\langle g|$ are the raising operators 
of the atom as defined in the previous section. 

In order to understand how the scheme can be made {\it near universal}, we 
rewrite (\ref{hamilto}) in terms of the $b_1$ and $b_2$-modes and redefined 
atomic states  $|e'_1\rangle$ and $|e'_2\rangle$ as
\begin{equation}
\label{newHamilto}
H_I=\hbar g \left[|e'_1\rangle\langle g|b_1+|e'_2\rangle\langle g|b_2+G'_1|e'_1\rangle\langle f|+G'_2|e'_2\rangle\langle f|\right] +\textrm{H.c.},
\end{equation}
where
\begin{equation}
\label{newterms}
|e'_1\rangle = \alpha |e_1\rangle +\beta |e_2\rangle~ ;~ |e'_2\rangle = \alpha^* |e_2\rangle - \beta^* |e_1\rangle ~;~ 
G'_1 = \alpha^* G_1+\beta^* G_2 ~;~ G'_2 = -\beta G_1+\alpha G_2.
\end{equation}
Clearly if we choose 
\begin{equation}
\label{univ_condn}
G_1/G_2 = - (\beta^*/\alpha^*),
\end{equation}
then the Hamiltonian in the new basis is like a single bias field acting on the
atomic transition. This analysis implies that we can deal with arbitrary input
states of the qubit by just choosing the bias field appropriately [Eq. (\ref{univ_condn})]. Note also the important property
\begin{equation}
\label{eq_denmat}
|e_1\rangle \langle e_1|+|e_2\rangle \langle e_2| = |e'_1\rangle \langle e'_1|+|e'_2\rangle \langle e'_2|, 
\end{equation}
so that the initial incoherent superposition remains an incoherent superposition
in the primed basis.
We next calculate the effect of the bias field on the fidelity of the system by
setting $G'_1=0$ in Eq. (\ref{newHamilto}). Let the eigenstates of $b_1^{\dagger}b_1$ and $b_2^{\dagger}b_2$ be denoted by $\widetilde{|n,m\rangle}$. From (\ref{newHamilto}) we find that the following states participate in time evolution:
\begin{equation}
|e'_1\rangle\widetilde{|1,0\rangle}~;~|e'_2\rangle\widetilde{|1,0\rangle}~;~|f\rangle\widetilde{|1,0\rangle}~;~|g\rangle\widetilde{|1,1\rangle}~;~|g\rangle\widetilde{|2,0\rangle}~;~|e'_1\rangle\widetilde{|0,1\rangle}.
\end{equation}
Note that the state $|e'_1\rangle\widetilde{|0,1\rangle}$ is produced by the
two-step process $|e'_2\rangle\widetilde{|1,0\rangle} \rightarrow |g\rangle\widetilde{|1,1\rangle} \rightarrow |e'_1\rangle \widetilde{|0,1\rangle}$.
We expand $|\Psi(t)\rangle$ in terms of these basis states and we obtain the 
following first order differential equations for the expansion amplitudes 
$\tilde{C}$'s:
\begin{eqnarray}
\dot{\tilde{C}}_{e_1}^{10}(t) & = & -\sq{2}i g \tilde{C}_g^{20}(t), \nonumber\\
\dot{\tilde{C}}_g^{20}(t) & = & -\sq{2}i g \tilde{C}_{e_1}^{10}(t),\nonumber\\ 
\dot{\tilde{C}}_{e_2}^{10}(t) & = & -i g \tilde{C}_g^{11}(t)-i gG'_2\tilde{C}_f^{10}(t), \nonumber\\
\dot{\tilde{C}}_f^{10}(t) & = & -igG'_2\tilde{C}_{e_2}^{10}(t), \\
\dot{\tilde{C}}_g^{11}(t) & = & -ig\tilde{C}_{e_1}^{01}(t)-ig \tilde{C}_{e_2}^{10}(t), \nonumber\\
\dot{\tilde{C}}_{e_1}^{01}(t) & = & -ig \tilde{C}_g^{11}(t)\nonumber.
\end{eqnarray}
Solving those equations subject to the initial conditions
\begin{equation}
\label{einitG}
\tilde{C}_{e_1}^{10}(0)=\frac{1}{\sq{2}}~~~~;~~~~\tilde{C}_{e_2}^{10}(0)=\frac{1}{\sq{2}}e^{i\theta},
\end{equation}
results
\begin{eqnarray}
\tilde{C}_g^{11}(t) & = & A\sin{\left(\frac{\Omega_1}{\sq{2}}gt\right)}+B\sin{\left(\frac{\Omega_2}{\sq{2}}gt\right)}, \nonumber\\
\tilde{C}_f^{10}(t) & = & \frac{1}{2G'_2}\left[(\Omega_1^2-4)A\sin{\left(\frac{\Omega_1}{\sq{2}}gt\right)}+(\Omega_2^2-4)B\sin{\left(\frac{\Omega_2}{\sq{2}}gt\right)}\right], \nonumber\\
\tilde{C}_{e_2}^{10}(t) & = & \frac{i}{2\sq{2}G'^2_2}\left[\Omega_1(\Omega_1^2-4)A\cos{\left(\frac{\Omega_1}{\sq{2}}gt\right)}+\Omega_2(\Omega_2^2-4)B\cos{\left(\frac{\Omega_2}{\sq{2}}gt\right)}\right],\\
\tilde{C}_{e_1}^{01}(t)&=&-\frac{i}{2\sq{2}G'^2_2}\left[\Omega_1(\Omega_1^2-2G'^2_2-4)A\cos{\left(\frac{\Omega_1}{\sq{2}}gt\right)}+\Omega_2(\Omega_2^2-2G'^2_2-4)B\cos{\left(\frac{\Omega_2}{\sq{2}}gt\right)}\right],\nonumber\\
\tilde{C}_{e_1}^{10}(t)&=&\frac{1}{\sq{2}}\cos{\left(\sq{2}gt\right)},\nonumber\\
\tilde{C}_g^{20}(t)&=&-\frac{i}{\sq{2}}\sin{\left(\sq{2}gt\right)},\nonumber
\end{eqnarray}
where 
\begin{displaymath}
 \Omega_1= \left(G'^2_2+2+\sq{G'^4_2+4}\right)^{\frac{1}{2}} ~~~~,~~~~\Omega_2=\left(G'^2_2+2-\sq{G'^4_2+4}\right)^{\frac{1}{2}},
\end{displaymath}
\begin{eqnarray}
\label{whoiswho2}
 A& = & \frac{1}{2}i e^{i\theta}\frac{(\Omega_2^2-2G'^2_2-4)}{\Omega_1\sq{G'^4_2+4}},\\
B & = & -\frac{1}{2}i e^{i\theta}\frac{(\Omega_1^2-2G'^2_2-4)}{\Omega_2\sq{G'^4_2+4}}.\nonumber
\end{eqnarray}
Using the relation (\ref{ereden}), we get the reduced density matrix of the 
field. The diagonal element of this density matrix in field basis 
$\widetilde{|k,l\rangle}$ gives the probability $\tilde{p}(k,l)$ that $k$ 
photons are in $b_1$-mode and $l$ photons are in $b_2$-mode and these are given
 by
\begin{eqnarray}
\tilde{p}(2,0)& = & |\tilde{C}_g^{20}(t)|^2 = \frac{1}{2}\sin^2{\left(\sq{2}gt\right)},\nonumber\\
\tilde{p}(1,1)& = & |\tilde{C}_g^{11}(t)|^2,\nonumber \\
\tilde{p}(1,0)& = & |\tilde{C}_f^{10}(t)|^2+|\tilde{C}_{e_2}^{10}(t)|^2+|\tilde{C}_{e_1}^{10}(t)|^2, \\
\tilde{p}(0,1)& = & |\tilde{C}_{e_1}^{01}(t)|^2.\nonumber
\end{eqnarray}
%The probabilities $\tilde{p}(2,0)$ and $\tilde{p}(1,1)$ after averaging over $\theta$ become
%\begin{equation}
%\tilde{p}_a(2,0)=\frac{1}{2}\sin^2{\left(\sq{2}gt\right)},\nonumber \\
%\tilde{p}_a(1,1)=\frac{1}{4(G^4+4)}\left[\frac{(\Omega_2^2-2G^2-4)^2}{\Omega_1^2}\sin^2{\left(\frac{\Omega_1}{\sq{2}}gt\right)}+\frac{(\Omega_1^2-2G^2-4)^2}{\Omega_2^2}\sin^2{\left(\frac{\Omega_2}{\sq{2}}gt\right)}\right].
%\end{equation}
Hence the fidelity $F$ takes the following form:%and $\tilde{F}$ take the following forms:
\begin{equation}
\label{efidelG}
%F(t)=\frac{\tilde{p}_a(2,0)+\frac{1}{2}\tilde{p}_a(1,1)}{\tilde{p}_a(2,0)+\tilde{p}_a(1,1)}~~~ ;~~~ 
F = 1 - \left[\tilde{p}_a(0,1)+\frac{1}{2}\tilde{p}_a(1,1)\right].
\end{equation}

\begin{figure}
\epsfxsize 10.8cm
{\center
\epsfbox{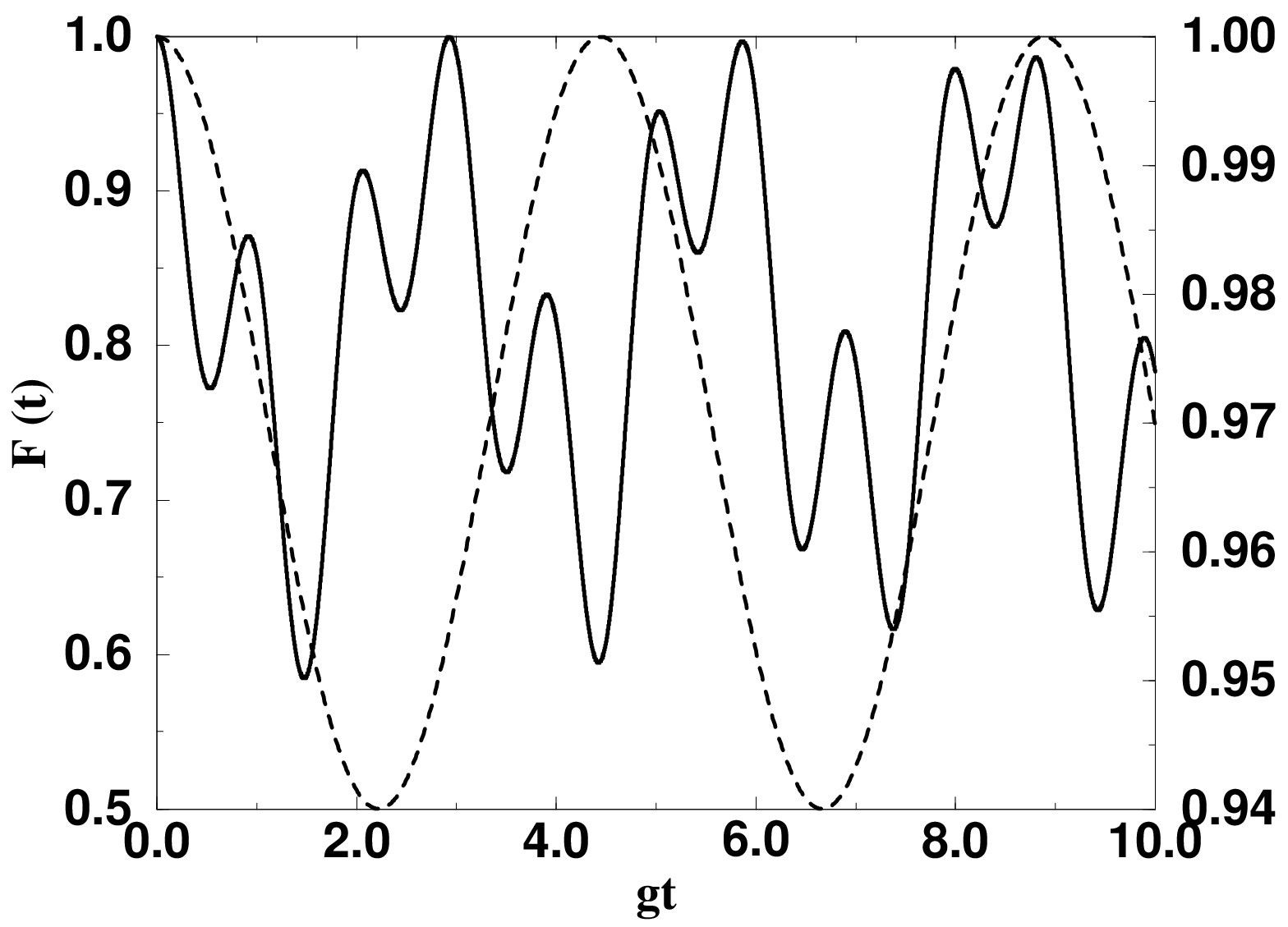}}
\caption{These figures show the time-dependence of the fidelity of a four-level
 atomic cloner comprising a single atom for no cycling field ($G'_1=G'_2=0$ ; 
dashed curve : tick levels are on left side) and in presence of external fields ($G'_1=0~,~G'_2=3$ ; solid 
curve : tick levels are on right side). It is obvious that the bias field improves the fidelity of cloning 
considerably.}  
\label{one}
\end{figure}

\begin{figure}
\epsfxsize 10cm
{\center
\epsfbox{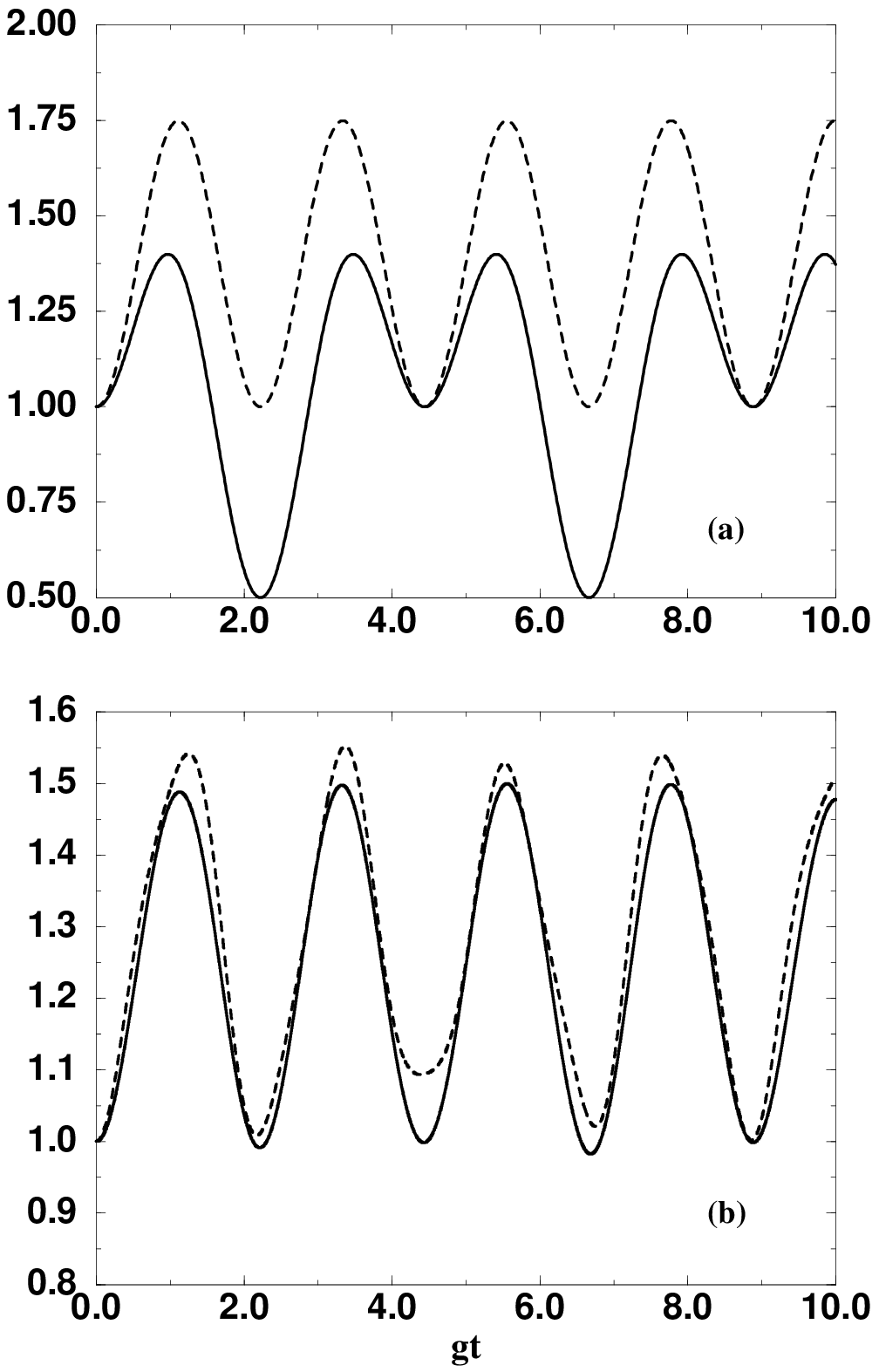}}
\caption{These figures display the time-dependence of average number of photons
 in ``right" mode, i.e., in $b_1$-mode ($N_{\mathrm{right}}$ ; solid curve) and
 of all photons ($N_{\mathrm{all}}$ ; dashed curve) in a single atom cloner 
under the conditions $G'_1=G'_2=0$ [Fig. (a)] and $G'_1=0~,~G'_2=3$ [Fig. (b)]. 
It is seen that for non-zero $G'_2$, 
$N_{\mathrm{right}}$ and $N_{\mathrm{all}}$ approach each other. We have seen 
that for $G'_2=8$, i.e., for faster cycling of the population, they are nearly 
equal for almost all times.}  
\label{avgN}
\end{figure}

We have plotted this as a function of time for $G'_2=0,3$ in Fig.\ \ref{one}. 
It is found that as $G'_2$ increases, the fidelity becomes  unity more often. 
Whenever both $\tilde{p}_a(1,1)$ and  $\tilde{p}_a(0,1)$ become zero, $F(t)$ becomes
 unity. In fact, by introducing the classical pump field, we cycle the atomic 
population in the state $|e'_2\rangle$ to the state $|f\rangle$ and back. This 
inhibits the spontaneous decay of the atom in the state $|e'_2\rangle$ to the 
ground state $|g\rangle$ irrespective of whether there is any photon or not
in ``right" mode. There are two time-scales of oscillation of $F(t)$. The 
faster small-amplitude oscillation is attributed to that of 
$[\tilde{p}_a(0,1)+(1/2)\tilde{p}_a(1,1)]$. This oscillation can be increased by
 $G'_2$ so that the atom effectively goes to the state  $|f\rangle$ very 
frequently. This means that $F(t)$ becomes close to unity more frequently. The 
effect of spontaneous 
emission from the state $|f\rangle$ is ignored assuming that the time scale for
this decay is much larger than that for Rabi oscillation between $|e'_2\rangle$ 
and $|f\rangle$-levels. In Fig.\ \ref{avgN} we display the mean number of 
``right" and of all photons. This figure also shows how the improvement in 
cloning is obtained by the use of a cycling or bias field.

\section{FIDELITY OF CLONING WITH TWO ATOMS}

\begin{figure}
\epsfxsize 11cm
\begin{center}
\epsfbox{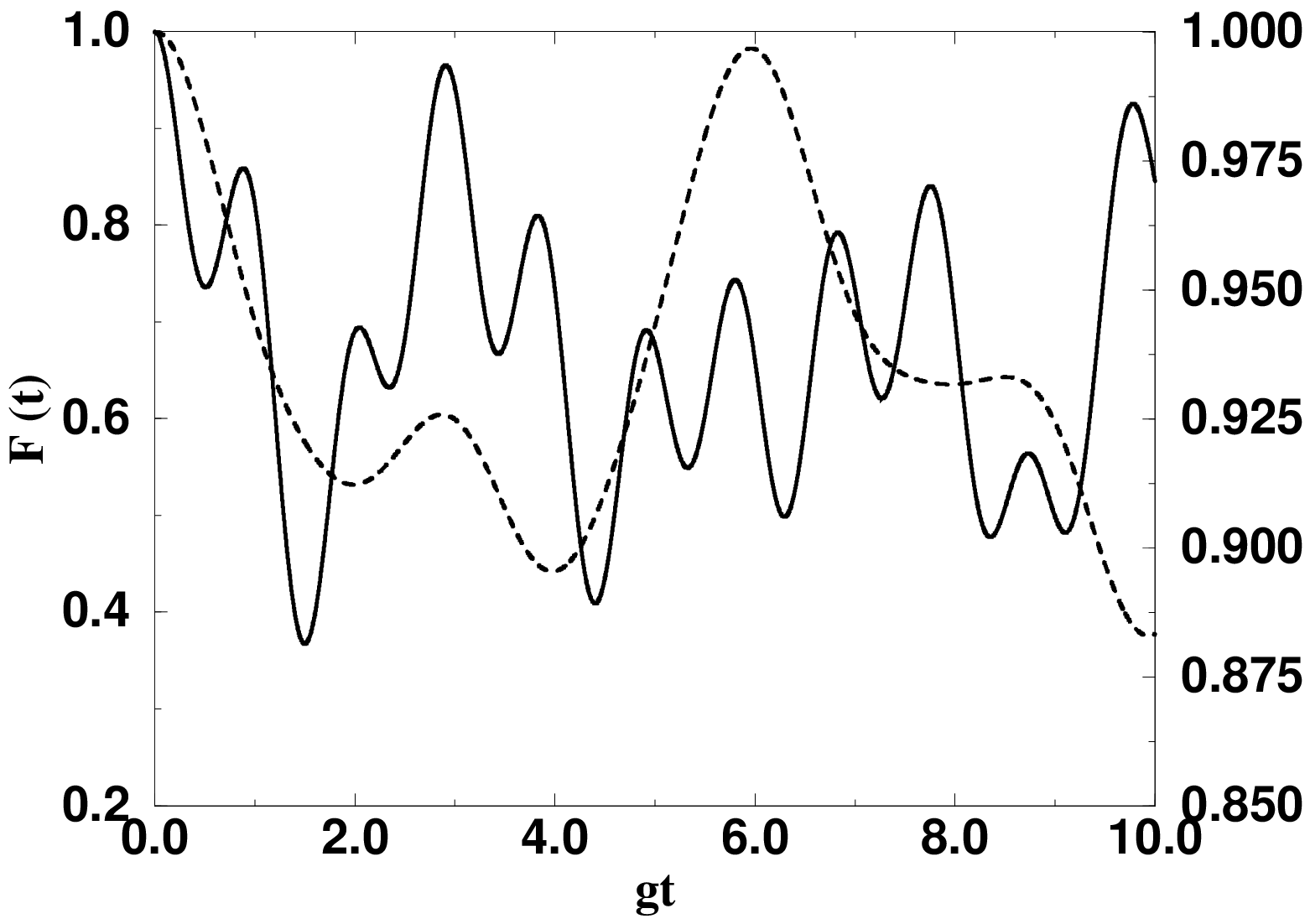}
\end{center}
\caption{The fidelity of a two-atom cloner is plotted as a function of time for
no cycling field ($G'_1=G'_2=0$ ; dashed curve : tick levels are on left 
side)  and in presence of the bias field ($G'_1=0~,~G'_2=3$ ; solid curve : tick
levels are on right side). Clearly the fidelity of cloning is improved by fast
cycling of atomic population by a classical bias field.}
\label{two}
\end{figure}

\begin{figure}
\epsfxsize 9.6cm
\begin{center}
\epsfbox{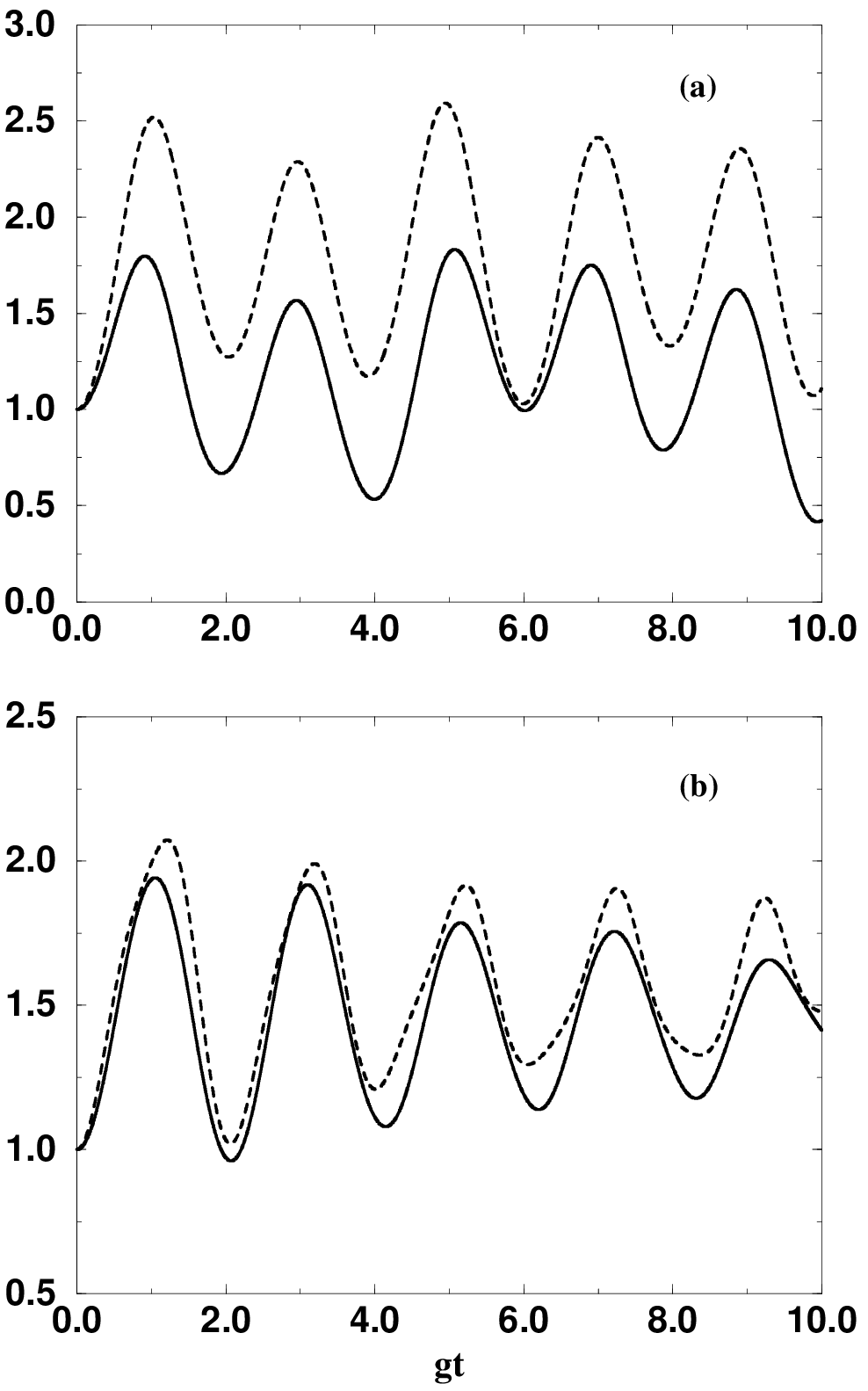}
\end{center}
\caption{The time-variation of average number of the photons in the $b_1$-mode 
($N_{\mathrm right}$ ; solid curve) and all photons ($N_{\mathrm all}$ ; dashed curve) in a two-atom cloner have been
displayed for $G'_1=G'_2=0$ [Fig. (a)] and $G'_1=0~,~G'_2=3$ [Fig. (b)]. This 
shows that $N_{\mathrm right}$ and $N_{\mathrm all}$ become closer for larger 
value of $G'_2$.}
\label{twoN}
\end{figure}

In this section, we consider a cloning machine consisting of two atoms and we
demonstrate improvement in the fidelity  
for a larger domain of times if we adopt the use of the cycling
 or bias field. We consider the case of two V-atoms. The interaction Hamiltonian
 is obtained by summing (\ref{newHamilto}) and a similar Hamiltonian involving 
the interaction of the another atom B with the fields. The initial state of the
 atomic system is 
\begin{equation}
\label{two_initial}
\rho = \prod_{\mu = {\mathrm A,B}} \left(|e_1\rangle_\mu\:_\mu\langle e_1|+|e_2\rangle_\mu\:_\mu\langle e_2|\right),
\end{equation}
and we assume single photon in the mode $b_1$. We assume that we work under the
 condition (\ref{univ_condn}) so that the Hamiltonian reduces to 
\begin{equation}
\label{two_hamilto}
H_I = \hbar g \sum_{\mu = {\mathrm A,B}}\left(|e'_1\rangle_\mu \:_\mu \langle g|b_1+|e'_2\rangle_\mu \:_\mu \langle g|b_2 + G'_2|e'_2\rangle_\mu\:_\mu\langle f|\right) + \textrm{H.c.}.
\end{equation}
In order to use the wavefunction picture we use the initial condition for the 
atom as 
\begin{equation}
\label{two_initial1}
\rho = \prod_{\mu = {\mathrm A,B}} |\Psi_\mu\rangle\langle \Psi_\mu| ~~~~~;~~~~~
|\Psi_\mu\rangle = \frac{1}{\sq{2}}\left(|e'_1\rangle_\mu + e^{i \theta_\mu}|e'_2\rangle_\mu\right),
\end{equation}
and average the final results over $\theta_{\mathrm A}$ and $\theta_{\mathrm B}$. For the two-atom 
case with bias field we have to use a large number of relevant basis states -- 
the size of the basis states increase very rapidly with increase in the number
of atoms.  The equations for the amplitudes are solved numerically using 
fifth-order Runge-Kutta method. From the numerical solutions we calculate the 
fidelity $F$.  It is clear from the Fig.\ \ref{two} that the cycling field 
makes the fidelity much higher for a very large range of times. Note that 
the fidelity of the two-atom cloner can be expressed as 
\begin{equation}
%F(t) = 1-\frac{\frac{1}{3}\tilde{p}_a(2,1)+\frac{2}{3}\tilde{p}_a(1,2)+\frac{1}{2}\tilde{p}_a(1,1)
%+\tilde{p}_a(0,2)}{\tilde{p}_a(2,0)+\tilde{p}_a(3,0)+\tilde{p}_a(2,1)+\tilde{p}_a(1,2)+\tilde{p}_a(1,1)+\tilde{p}_a(0,2)}.
F(t) = 1-\left[\frac{1}{3}\tilde{p}_a(2,1)+\frac{2}{3}\tilde{p}_a(1,2)+\frac{1}{2}\tilde{p}_a(1,1)+\tilde{p}_a(0,1)+\tilde{p}_a(0,2)\right].
\end{equation}
Obviously, it becomes unity only if the probabilities of spontaneous emission in
$b_2$-mode (both in presence and in absence of photons in $b_1$-mode) are zero.
Then all the photons present in the cavity would be in $b_1$-mode. However due 
to complex nature of time-dependence of $\tilde{p}_a(k,l)$'s, one does not find 
any periodicity in the variation of $F(t)$.
% On comparing Figs.\ \ref{one} and 
%\ \ref{two}, we find that the fidelity of cloning is higher for the case of a 
%cloner consisting of two atoms than for the case of a cloner consisting of a 
%single atom. 
The average number of ``right" photons and of all photons in the
evolved state of the two-atom cloner have been plotted as functions of time 
in the Fig.\ \ref{twoN}. A comparison of Figs.\ (\ref{twoN}a) and (\ref{twoN}b)
shows tremendous improvement in cloning due to the cycling field.

\section{AVERAGE FIDELITY OF CLONING FOR A FIXED BIASING FIELD}

\begin{figure}
\epsfxsize 10cm
{\center
\epsfbox{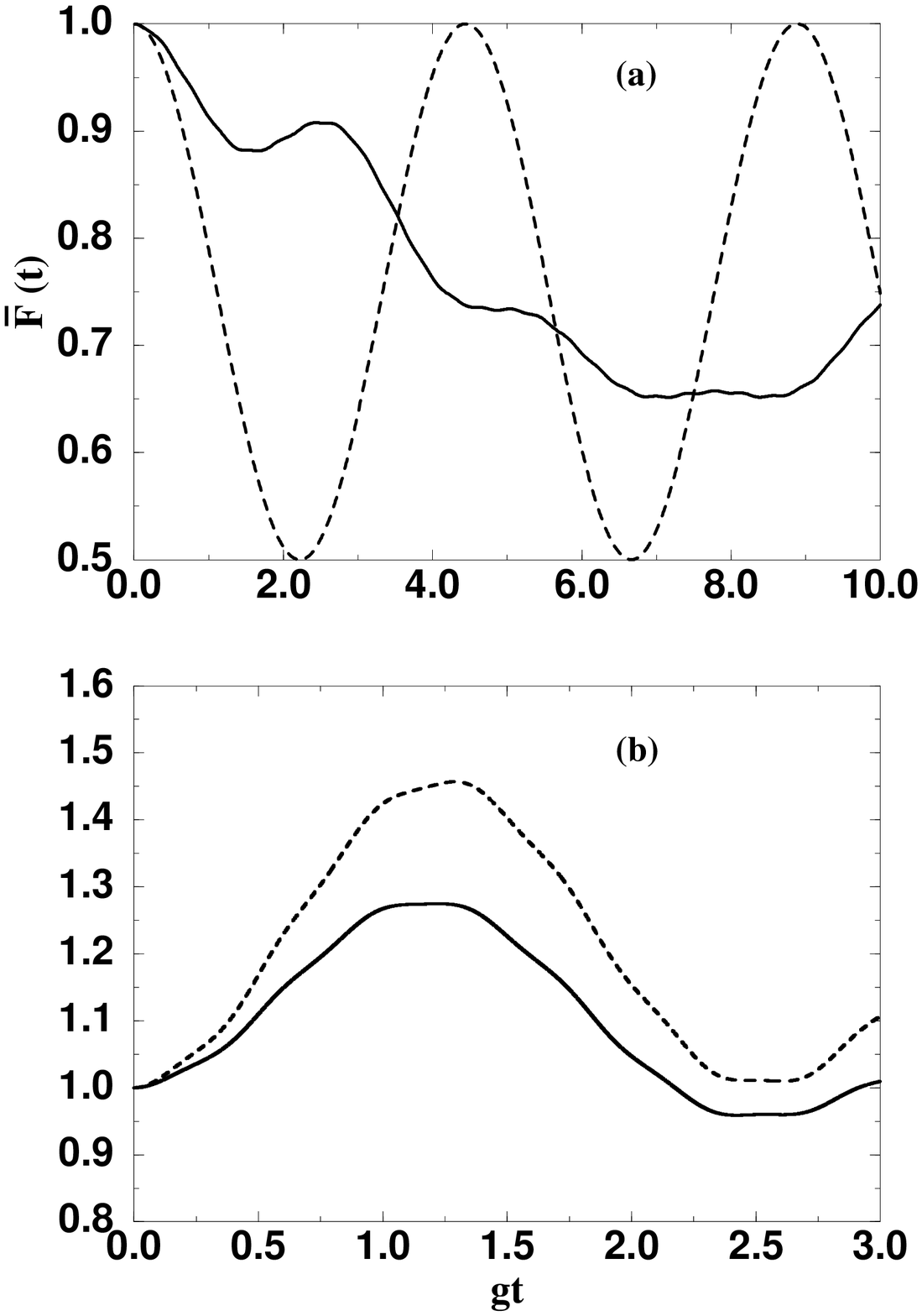}}
\caption{(a) The fidelity of a single atom cloner averaged over all states of 
the input qubit is plotted as a function of time for the cases $G'_1=G'_2=0$ 
(dashed curve) and $G'_1=0~,~G'_2=8$ (solid curve). (b) The time-variation of 
number of ``right" photons (solid curve) and all photons (dashed curve), 
averaged over all states of the input qubit are shown for some external field 
parameters $G'_1=0~,~G'_2=8$.}  
\label{FNavgAB}
\end{figure}

In the previous sections we had discussed the possibility of improvement in
fidelity by changing the bias field as one changes the input state of the qubit.
The question arises what is the fidelity of cloning for a fixed bias field. In
such a case we have to work with the average of fidelity over all the input
states of the qubit. To be precise let us consider the input state of the qubit
as given by Eq. (\ref{einitgen1}). We calculate the probability $\tilde{p}(k,l)$
 as defined by (\ref{eprobb}). We next average this probability over all values
of $\alpha$ and $\beta$: $\alpha = \cos{(\chi/2)}, \beta = \sin{(\chi/2)}e^{i\varphi}$, where $0 \leq \chi \leq \pi ~,~ 0 \leq \varphi \leq 2\pi$. Using average
probabilities $\overline{p}(k,l)$ we obtain the average fidelity for all the 
states of the input qubit. The calculations are lengthy and we present results
in the Fig.\ \ref{FNavgAB}a. Note that for no bias, the fidelity is periodic in time whereas in
presence of the bias field it is quasiperiodic. We conclude from this figure 
that there is considerable {\it improvement} in fidelity upto times of order 
$gt \sim 2$. This is also reflected by the behavior of the mean photon number
in the ``right" mode (Fig.\ \ref{FNavgAB}b).

\section{CONCLUSIONS}

In conclusion, we have proposed how the fidelity of a V-system-based quantum 
cloner can be improved by inhibiting the spontaneous emission effects on the 
unwanted transition. We showed that this can be done by applying a classical 
coherent field, which cycles the atom in the state $|e_2\rangle$ to some other 
metastable state. The fidelity remains close to unity over large intervals of 
time. Furthermore the fidelity of a cloning machine based on two atoms inside 
the cavity is much better provided we continue to use a cycling field. However 
we must add that in
order to maintain the universality of the scheme we have to choose a cycling
field which depends on the state of the input qubit. Otherwise for a fixed 
cycling field one has to be content with an average fidelity which for a range
of interaction times is also found to be much better compared to the one in the
absence of the cycling field.

GSA thanks P. L. Knight and A. Zeilinger for discussions on this subject.

\begin{thebibliography}{}
\bibitem{space}
This means that all the atoms are in equivalent positions relative to the cavity
mode. This condition can be achieved by trapping the atoms inside the cavity, so that the 
field interacting with the atom is the same for all atoms. This kind of trapping of
atoms has been reported recently  by J. Ye,
 D. W. Vernooy, and H. J. Kimble, {\bf quant-ph}/9908007.
\end{thebibliography}
\end{document}